\documentstyle [12pt,amsfonts] {article}
\input epsf

\newcommand{\beq}{\begin{equation}}
\newcommand{\eeq}{\end{equation}}
\newcommand{\beqs}{\begin{eqnarray}}
\newcommand{\eeqs}{\end{eqnarray}}

\catcode`@=11
\@addtoreset{equation}{section}
\@addtoreset{equation}{subsection}
\def\theequation{\ifnum\value{section}=0 \arabic{equation}\ignorespaces
\else \ifnum\value{section}=-1 A.\arabic{equation}\ignorespaces
\else \ifnum\value{subsection}=0 \thesection.\arabic{equation}\ignorespaces
\else \thesection.\arabic{subsection}.\arabic{equation}\ignorespaces
                           \fi
                      \fi
                 \fi}
\catcode`@=12

\begin{document}

\def\thefootnote{\fnsymbol{footnote}}

\baselineskip 6.0mm

\vspace{4mm}

\begin{center}

{\Large \bf Zeros of Jones Polynomials for Families of Knots and Links} 

\vspace{8mm}

\setcounter{footnote}{0}
S.-C. Chang\footnote{email: shu-chiuan.chang@sunysb.edu}, 
\setcounter{footnote}{6}
R. Shrock\footnote{email: robert.shrock@sunysb.edu}

\vspace{6mm}

C. N. Yang Institute for Theoretical Physics  \\
State University of New York       \\
Stony Brook, N. Y. 11794-3840  \\

\vspace{6mm}

{\bf Abstract}
\end{center}

We calculate Jones polynomials $V_L(t)$ for several families of alternating
knots and links by computing the Tutte polynomials $T(G,x,y)$ for the 
associated graphs $G$ and then obtaining $V_L(t)$ as a special case of the
Tutte polynomial.  For each of these families we determine the zeros of the 
Jones polynomial, including the accumulation set in the limit of infinitely 
many crossings.  A discussion is also given of the calculation of Jones
polynomials for non-alternating links. 

\vspace{16mm}

\pagestyle{empty}
\newpage

\pagestyle{plain}
\pagenumbering{arabic}
\renewcommand{\thefootnote}{\arabic{footnote}}
\setcounter{footnote}{0}

\section{Introduction}

Knots and links are of interest in mathematics and because of their connection
to Yang-Baxter relations and solvable models \cite{rolfsen}-\cite{jones89}.
Here a link is defined as an embedding of $n_c$ circles $S^1$ (each forming a
component of the link) in ${\mathbb R}^3$, and a knot is the special case
$n_c=1$ of a link involving one such circle.  We shall generally use the
notation $L$ and $K$ for links and knots and $L$ when referring to both.  A
longstanding problem in mathematics has been to reduce the classification of
knots and links to an algebraic formulation. Various polynomials have been
constructed in an effort to achieve an algebraic classification of knots.  Here
we shall focus on the Jones polynomial $V_L(t)$ \cite{jones85,jones87}.
Technically, this is not, in general, a polynomial, since it can involve
non-integral powers of its variable $t$ and, even in cases where it only
involves integral powers, these may be negative.  However, $V_L(t)$ can always
be written as a polynomial in $t$ multiplied by a monomial prefactor.  With
this qualification, we shall follow common usage in referring to $V_L(t)$ as a
polynomial. Lists of knots and links are given with standard ordering and
labelling in, e.g., \cite{rolfsen}, and we follow the labelling conventions in
this reference.  The usual presentation of knots and links involves a
projection from the embedding space ${\mathbb R}^3$ to ${\mathbb R}^2$ with
resultant apparent crossings of the curves on the one or more embedded $S^1$'s.
According to standard labelling, $5_2 \equiv 5^1_2$ denotes the second knot
with five crossings in the list, $6^3_2$ denotes the second link containing
three intertwined components with six crossings, etc.  We recall that an
alternating link is defined as one in which, as one travels along each of the
$n_c$ embedded circles $S^1$, one traverses a crossing in an overlying manner,
then underlying, then overlying, and so forth.

Here we shall calculate Jones polynomials for certain families of knots and,
more generally, links with arbitrarily many crossings and analyze certain
properties in the limit of infinitely many crossings.  An interesting
pioneering study of this type has been carried out recently by Wu and Wang
\cite{ww}.  We shall concentrate on alternating links but shall also briefly
discuss the non-alternating case.  Let us denote the member of a generic family
as $L_r$, meaning that this has $r$ crossings. Other information about the
links will be indicated as necessary below but is suppressed here.  We
introduce the notation for the formal limit $\lim_{r \to \infty} L_r = \{L\}$.
It is also useful to investigate the zeros of the Jones polynomial for classes
of links or knots in the complex $t$ plane and their accumulation set ${\cal
B}$ in the limit where the number $r$ of crossings goes to infinity.

There are several motivations for this study.  It is worthwhile to have new
exact calculations of Jones polynomials for various families of links;
furthermore, the fact that these apply for arbitrarily many crossings enables
one to investigate the limit where the number of crossings goes to infinity.
As noted above, knots and links are also relevant to physics because of their
relation to Yang-Baxter equations.  These equations are conditions for (i)
transfer matrices of exactly soluble models in statistical mechanics, and (ii)
equivalently, $S$-matrix elements of relativistic particle scattering in one
space and one time dimension, so that these $S$ matrix elements factorize,
i.e., the $S$ matrix an $n \to n$ scattering process factorizes into a product
of $S$ matrix elements for simple $2 \to 2$ scatterings, without particle
production.  Indeed, there are well-known connections between Yang-Baxter
relations, the braid group, and Temperley-Lieb algebras (e.g.,
\cite{yg}-\cite{wu}).  As we shall discuss next, there are also useful
relations between the Jones polynomial and the Tutte polynomial, or
equivalently, the partition function of the Potts model, a valuable model of
phase transitions and critical phenomena.

\section{Some General Results} 

\subsection{Connection between Jones and Tutte Polynomials} 

Our method for calculating the Jones polynomials $V_L(t)$ for a given family of
alternating links is to compute the Tutte polynomial $T(G,x,y)$ for a certain
associated graph $G$ and then use the fact that $V_L(t)$ can be obtained as a
special case of the Tutte polynomial for $x=t$ and $y=-1/t$
\cite{jones89,wu,boll}.   To explain this method, we first discuss some
background. For a given knot, draw the associated shaded-region diagram. There
are two such diagrams.  To see this, consider a given crossing, say in the form
of an $\times$; in one shading, the regions to the left and right of the
$\times$ are light and those above and below the $\times$ are dark.  In the
other shading, the regions to the left and right are dark while those above and
below are light. These two possible shadings are labelled as plus and minus
with the following convention: as one exits a given crossing on the curve that
is overlying, if the region on the left (right) is shaded, assign $+$ ($-$) to
this crossing.  For a given knot or link $L$, the numbers of the $\pm$
crossings are denoted $r_{\pm}(L)$.  It is easily seen that for an alternating
knot, if one crossing on the shaded-region diagram is $+$, then all are, and
similarly if one crossing is $-$, then all are.  We denote the corresponding
shaded region diagrams as $D_+(L)$ and $D_-(L)$.  We denote the numbers of dark
($d$) and light ($\ell$) regions in the shaded region diagram $n_d(D_\pm)$ and
$n_\ell(D_\pm)$.  Clearly, $n_d(D_+)=n_\ell(D_-)$ and
$n_\ell(D_+)=n_d(D_-)$. From these shaded-region diagrams $D_\pm(L)$, one then
constructs the associated graphs $G_+(L)$ and $G_-(L)$ by assigning vertices to
the shaded regions and edges (bonds) connecting these vertices.  We recall that
a graph $G=(V,E)$ is defined by its vertex and edge sets $V$ and $E$.  Strictly
speaking, the objects that we are dealing with here are pseudographs, since
they can have multiple edges and loops, but we shall continue to call them
simply graphs. Let the number of vertices and edges of the graph $G$ be denoted
$n(G)=|V|$ and $e(G)=|E|$. Clearly, the total number of crossings is equal to
the number of edges of the associated graph:
\beq
r(L)=e(G_+(L))=e(G_-(L))
\label{ncr}
\eeq
and the respective numbers of dark (light) regions in the $D_+$ ($D_-$) 
region diagrams is equal to the number of vertices in the associated graphs:
\beq
n_d(D_+)=n(G_+(L)), \quad n_d(D_-)=n(G_-(L)) \ .
\label{ndrel}
\eeq
Furthermore, 
\beq
G_+(L)=[G_-(L)]^*
\label{gdual}
\eeq
i.e., $G_-(L)$ is the (planar) dual of $G_+(L)$.  (Here, from a planar graph
$G$, the (planar) dual $G^*$ is constructed by assigning to each face of $G$ a
vertex and connecting these vertices by edges traversing the edges of $G$.)
Another standard definition that we will need is the writhe of the knot or link
$L$, denoted $w(L)$.  For this we first define another type of sign $s=\pm$ 
for each crossing in a link.  For each circle on the link, we consider an
orientation, which can be regarded as a direction of motion along the
circle. Next, we represent a given crossing as two arrows, one pointing to the
upper left and one to the upper right.  If the lower-right to upper-left arrow
is over (under) the lower-left to upper-right arrow, the sign $s$ is $-$ ($+$).
Denote the total number of crossings with $s=+$ ($s=-$) as $N_{s_+}$ (
$N_{s_-}$), respectively. Then the writhe is
\beq
w(L)=(+1)N_{s_{+}}+(-1)N_{s_{-}}
\label{wl}
\eeq
To each knot or link $K$ or $L$, there corresponds another, which we
label as $R(K)$ and $R(L)$ obtained by reversing all crossings, i.e. replacing
an over-under crossing by an under-over crossing, and vice versa.  A basic
property of the Jones polynomial is that
\beq
V_L(t)=V_{R(L)}(t^{-1}) \ .
\label{vlvrl}
\eeq

We next proceed to the relation between the Jones polynomial $V_L(t)$ and the 
Tutte polynomial of the associated graph $G_+(L)$ which applies for the case in
which $L$ is an alternating link or knot.  The Tutte polynomial $T(G,x,y)$ of a
graph $G$ (which may include multiple edges and loops) is 
\cite{tutte1}-\cite{tutte3}
\beq
T(G,x,y)=\sum_{G^\prime \subseteq G} (x-1)^{k(G^\prime)-k(G)}
(y-1)^{c(G^\prime)}
\label{tuttepol}
\eeq
where $G^\prime$ is a spanning subgraph of $G$, i.e., $G^\prime$ has the same
vertex set and a subset of the edge set, of $G$: $G^\prime = (V,E^\prime)$ with
$E^\prime \subseteq E$.  In eq. (\ref{tuttepol}), $k(G^\prime)$,
$e(G^\prime)=|E^\prime|$, and $c(G^\prime)$ denote the number of components,
edges, and (linearly independent) circuits of $G^\prime$, with the usual
relation
\beq
c(G^\prime) = e(G^\prime)+k(G^\prime)-n(G^\prime)
\label{ceq}
\eeq
where, as above, $n(G^\prime)$ is the number of vertices of $G^\prime$. 
The Tutte polynomials of a planar graph $G$ and its (planar) dual $G^*$ are
related according to
\beq
T(G,x,y)=T(G^*,y,x) \ .
\label{ttdual}
\eeq

The relation between the Jones polynomial $V_L(t)$ and the Tutte polynomial 
of the associated graph $G_+(L)$ for an alternating link or knot is then 
\beq
V_L(t)=(-1)^w t^{(n_\ell(G_+)-n_d(G_+)+3w)/4} T(G_+(L),-t,-1/t) \ .
\label{vt}
\eeq
Note that for a link involving an odd number $n_c$ of components (circles) and
hence, in particular, for a knot, $n_\ell(G_+)-n_d(G_+)+3w$ is always a
multiple of 4, so that $V_L(t)$ is a multinomial, i.e., Laurent polynomial, in
$t$. For a link consisting of an even number $n_c \ge 2$ of components 
$n_\ell(G_+)-n_d(G_+)+3w = 2$ mod 4, so that $V_L(t)$ is of the form
$V_L(t)=t^qPoly_L(t)$, where $q$ is a positive or negative half-odd-integer.

\subsection{Some Further Properties}

A basic property of the Jones polynomial $V_L(t)$ for a link $L$ is that
\cite{jones87} 
\beq
V_L(e^{2\pi i/3})=(-1)^{n_c}
\label{vl2p3}
\eeq
where, as before, $n_c$ denotes the number of $S^1$'s in the link.  Since in
general, $V_L(t)$ (aside from its prefactor) may be a high-order polynomial in
$t$, this suggests that $V_L(t)$ may have a complex-conjugate pair of zeros 
near to $t=e^{\pm 2 \pi i/3}$.  Indeed, for the special case $n_c=1$, i.e., a
knot, one has \cite{jones87}
\beq
V_K(t)=1-(1-t)(1-t^3)W_K(t)
\label{vw}
\eeq
where $W_K(t)$ is a Laurent polynomial in $t$.  As shown in \cite{ww}, it
follows that $V_K(t)$ has a complex-conjugate pair of zeros that approach the
points
\beq
t=e^{\pm 2 \pi i/3}
\label{tpoints}
\eeq
as $r(K) \to \infty$.

Two other basic properties concern the highest and lowest powers of $t$ in
$V_L(t)$ for an alternating knot or link.  To determine these, we first derive
the values of the respective maximal degrees of the Tutte polynomial in its 
variables $x$ and $y$: 
\beq
{\rm max}({\rm deg}_x[T(G,x,y)]) = n(G)-1
\label{maxdegx}
\eeq
and
\beq
{\rm max}({\rm deg}_y[T(G,x,y)]) = c(G) \ .
\label{maxdegy}
\eeq
To prove (\ref{maxdegx}) we maximize the power $k(G^\prime)-k(G)$ in the
definition (\ref{tuttepol}).  This maximum power is realized for $G^\prime =
(V,\emptyset)$, i.e., the case where the spanning subgraph has no edges.
For the case of interest here, where $G$ itself is connected, this power then
takes on the maximal value $n(G)-1$.  To prove (\ref{maxdegy}), we want to
maximize the number of circuits in $G^\prime$.  Clearly, 
$c(G^\prime) \le c(G)$, so we achieve this maximum number of circuits by taking
$G^\prime = G$.  

Then, combining (\ref{maxdegx}) and (\ref{maxdegy}) with (\ref{vt}), we have,
for an alternating link (indicating also the sign of $t$)
\beq
{\rm max}({\rm deg}[V_L(t)]) =
 (-1)^{w(L)+n(G_+(L))-1} t^{n(G_+(L))-1 + (n_\ell(G_+(L))-n_d(G_+(L))+3w(L))/4}
\label{hipowvl}
\eeq
\beq
{\rm min}({\rm deg}[V_L(t)] =  (-1)^{w(L)-c(G_+(L))}t^{-c(G_+(L))+
(n_\ell(G_+(L))-n_d(G_+(L))+3w(L))/4}
\label{lopowvl}
\eeq
 From these results together with the
relation (\ref{ceq}), it follows that 
\beqs
{\rm max}({\rm deg}[V_L(t)])-
{\rm min}({\rm deg}[V_L(t)]) & = & e(G_+(L)) \cr\cr
                                          & = & r(L) \ .
\label{degdif}
\eeqs
We can simplify the sign in Eq. (\ref{hipowvl}). Since $(-1)^{w(L)} =
(-1)^{r(L)} = (-1)^{e(G_+(L))}$, we have
\beq
(-1)^{w(L)+n(G_+(L))-1}=(-1)^{n_d(D_-(L))-1}
\label{mineq}
\eeq
which makes use of the Euler relation for a planar graph
$n(G)-e(G)+f(G)=2$,
where $f(G)=v(G^*)$ is the number of faces of $G$, which is also equal to the
number of vertices of the planar dual, $G^*$.

\subsection{Limit of Links with Infinitely Many Crossings} 

We next define a function 
\beq
U(\{L\},t) = \lim_{n(G_+(L)) \to \infty} [(-1)^{n_d(D_-(L))-1}V_{L}(t)]^
{1/n(G_+(L))} \ .
\label{udef}
\eeq
With the multiplicative factor incorporated as indicated, the coefficient of
the maximal power of $t$ is positive and hence the resultant function is real
and positive for large real positive $t$.  Consequently, among the $n$
different $1/n$'th roots, one can, with no ambiguity, pick the real positive
one, and this is understood in eq. (\ref{udef}).  The largest region in the
complex $t$ plane to which one can analytically continue the function 
$U(\{L\},t)$ we shall call region $R_1$.   

We remark that there is a subtlety in the definition (\ref{udef}) due to the
fact that at certain special points $t=t_s$ one can encounter the 
noncommutativity 
\beq
\lim_{t \to t_s} \lim_{n \to \infty} [V_{L}(t)]^{1/n} \ne 
\lim_{n \to \infty} \lim_{t \to t_s}  [V_{L}(t)]^{1/n}
\label{unoncom}
\eeq
where $n \equiv n(G_+(L))$.  Denote $U_{tn}$ and $U_{nt}$ as the functions
defined by the different order of limits on the left and right-hand sides of
(\ref{unoncom}).  Here we make the choice
\beq
U(\{L\},t) \equiv U_{tn}(\{L\},t)
\label{ulchoice}
\eeq
since this has the advantage of removing certain isolated discontinuities that
are present in $U_{nt}$.

The definition (\ref{udef}) is analogous to
the definition of the exponent of the reduced free energy $f$ in statistical
mechanics according to 
\beq
e^f = \lim_{n \to \infty} Z^{1/n}
\label{ef}
\eeq
where $Z$ is the partition function.  Again, in this definition one implicitly
takes the real positive $1/n$'th root, and there is no ambiguity since for 
physical values of temperature and external magnetic field, $Z$ is real and
positive.  Another analogy is the ground state degeneracy $W$ per site for the
Potts antiferromagnet on a graph $G$, which can be obtained from the chromatic
polynomial $P(G,q)$, counting the number of ways of coloring the vertices of a
graph $G$ subject to the constraint that no two adjacent vertices have the same
color.  Denoting the formal limit as $n \to \infty$ of a family of graphs $G$
as $\{G\}$, we define 
\beq
W(\{G\},q)=\lim_{n \to \infty}P(G,q)^{1/n} \ .
\label{wdef}
\eeq
In \cite{w} it was noted that at certain special points $q_s$ (typically
$q_s=0,1,.., \chi(G)$, where $\chi(G)$ is the chromatic number of the graph
$G$, i.e. the minimum number of colors necessary for a proper vertex coloring
of $G$), one has the noncommutativity of limits
\beq
\lim_{q \to q_s} \lim_{n \to \infty} P(G,q)^{1/n} \ne \lim_{n \to
\infty} \lim_{q \to q_s}P(G,q)^{1/n}
\label{wnoncom}
\eeq
and hence it is necessary to specify the order of the limits in the definition
of $W(\{G\},q_s)$ \cite{w}. Denoting $W_{qn}$ and $W_{nq}$ as the functions
defined by the different order of limits on the left and right-hand sides of
(\ref{wnoncom}), the choice $W \equiv W_{qn}$ was made in \cite{w} and
subsequent our works since this has the advantage of removing certain isolated
discontinuities that are present in $W_{nq}$.  For a given family of graphs
$G$, and for sufficiently large real positive $q$, $P(G,q)$ is real and
positive, so again one can choose, without any ambiguity, the real positive
$1/n$'th root in (\ref{wdef}).  In \cite{w} the largest region in the complex
$q$ plane to which one can analytically continue the function $W(\{G\},q)$ from
the real positive axis for large $q$ was denoted as $R_1$.  Just as
$W(\{G\},q)$ is analytic in certain regions of the complex $q$ plane,
$U(\{L\},t)$ is analytic in certain regions of the complex $t$ plane.  And just
as was true for $W$, in regions other than $R_1$, $V_L(t)$ can be negative or
complex, hence there is no canonical choice of the phase to take in evaluating
the $1/n$'th root in (\ref{udef}), and only the magnitude $|U(\{L\},t)|$ can be
determined unambiguously.

The continuous locus where $U(\{L\},t)$ is nonanalytic, called ${\cal B}$,
arises as the continuous accumulation set of the zeros of $V_L(t)$.  A
pioneering discussion of this type of accumulation of zeros to form
non-analytic boundaries for a model in statistical physics was given by Yang
and Lee \cite{yl,ly} for the zeros of the Ising model partition function as a
function of the complex variable $\mu=e^{-2\beta H}$, where
$\beta=(k_BT)^{-1}$, $T$ is the temperature, and $H$ is the external magnetic
field.  A second example is the zeros of a spin model partition function in the
complex variable $a^K$, where $K=\beta J$ and $J$ is a spin-spin exchange
coupling \cite{fisher}.  A more complicated example is provided by the study of
the zeros and their accumulation set in the ${\mathbb C}^2$ space defined by
$(a,\mu)$ \cite{ih,yy}.

\subsection{Connection with Potts Model Partition Function}

It is useful to observe that the Tutte polynomial of a graph is essentially
equivalent to the partition function of the Potts model on this graph (for a
review of the Potts model, see \cite{wurev}).  We recall that on $G$ (often
taken to be a regular lattice in physics) at temperature $T$ the Potts model is
defined by the partition function
\beq
Z(G,q,v) = \sum_{ \{ \sigma_n \} } e^{-\beta {\cal H}}
\label{zfun}
\eeq
with the (zero-field) Hamiltonian
\beq
{\cal H} = -J \sum_{\langle i j \rangle} \delta_{\sigma_i \sigma_j}
\label{ham}
\eeq
where $\sigma_i=1,...,q$ are the spin variables on each vertex $i \in G$;
$\beta = (k_BT)^{-1}$; and $\langle i j \rangle$ denotes pairs of adjacent
vertices, and again with $K=\beta J$, 
\beq
v = e^K-1 \ .
\label{vdef}
\eeq
This function $Z(G,q,v)$ can be written as the sum \cite{kf}
\beqs
Z(G,q,v) & = & \sum_{G^\prime \subseteq G} q^{k(G^\prime)}v^{e(G^\prime)}
\label{cluster} \cr\cr\cr
& = & \sum_{r=k(G)}^{n(G)}\sum_{s=0}^{e(G)}z_{rs} q^r v^s \ .
\label{zpol}
\eeqs
This formula allows one to generalize the definition of the Potts model from $q
\in {\mathbb Z}_+$ to $q \in {\mathbb C}$ and from the real physical range 
$-1 \le v < \infty$ to $v \in {\mathbb C}$.  Now let
\beq
x=1+\frac{q}{v}
\label{xdef}
\eeq
and
\beq
y=v+1
\label{ydef}
\eeq
so that
\beq
q=(x-1)(y-1) \ .
\label{qxy}
\eeq
Then the Tutte polynomial and Potts model are related according to 
\beq
Z(G,q,v)=(x-1)^{k(G)}(y-1)^{n(G)}T(G,x,y) \ .
\label{ztutte}
\eeq
Another relation concerns the chromatic polynomial $P(G,q)$, which counts the
number of ways of coloring the vertices of $G$ subject to the condition that no
two adjacent vertices have the same color \cite{rrev,rtrev}.  This is identical
to the partition function of the Potts antiferromagnet at $T=0$, i.e., $v=-1$:
\beq
P(G,q)=Z(G,q,-1)
\label{zp}
\eeq
or, equivalently, 
\beq
P(G,q)=(-q)^{k(G)}(-1)^{n(G)}T(G,1-q,0) \ .
\label{pt}
\eeq
Thus, via the connection to the Tutte polynomial or Potts model partition
function, one sees that the study of the zeros of the Jones polynomial is
related to studies of complex-temperature zeros of Ising and Potts models
\cite{fisher}, \cite{ipz}-\cite{p2}, to studies of zeros in complex $q$ of
chromatic polynomials (e.g., \cite{bds}-\cite{s4},\cite{w}), and the studies of
zeros of Potts model partition functions in the ${\cal C}^2$ space spanned by
$(q,v)$ \cite{a}-\cite{ta}, \cite{sbounds}. 

It will be convenient to recall some definitions from graph theory.  The
circuit graph $C_n$ is the graph of $n$ vertices connected with edges to form a
circle.  The (planar) dual of this graph is called the fat link, $FL_n$, and
consists of two vertices with $n$ edges connecting them.  The join $G+H$ of two
graphs $G$ and $H$ is the graph formed by connecting each vertex of $G$ to all
of the vertices of $H$.  The wheel graph $(Wh)_n$ is the join of a single
vertex with the circuit graph $C_{n-1}$.  Note that the wheel graph is
self-dual: $(Wh)_n^*=(Wh)_n$. The degree of a vertex is the number
of edges that connect to it.  A homeomorphic expansion (=
inflation) of a graph $G$ is obtained by inserting one or more degree-2
vertices to edges of $G$.  Given a graph, the homeomorphic expansion of this
graph is another graph.  The inverse operation, homeomorphic reduction,
consists in removing one or more degree-2 vertices from a graph.  This can take
one out of the class of proper graphs by resulting in a multigraph; for
example, the homeomorphic reduction of $C_3$ is the multigraph $C_2$.

\section{A Structural Theorem for $V_L(t)$}

We next derive a basic theorem which will be important for our further
analysis.  It was proved in \cite{a} that the Tutte polynomial for a recursive
graph consisting of $m$ repetitions of some basic subgraph can be written in
the form 
\beq
T(G,x,y) = \sum_{j=1}^{N_{T,G,\lambda}} c_{G,j}(\lambda_{G,j})^m 
\label{tgsum}
\eeq
where the coefficients $c_{G,j}$ and the terms $\lambda_{G,j}$ depend on the
type of graph but not on $m$.  It follows from this and the relation (\ref{vt})
for alternating links that the Jones polynomial for links $L$ whose associated
graphs $G_\pm (L)$ are recursive has the same type of form,
\beq
V_L(t) = \sum_{j=1}^{N_{L,\lambda}} c_{L,j}(\lambda_{L,j})^m
\label{vlsum}
\eeq
where 
\beq
N_{L,\lambda} \le N_{T,G_+(L),\lambda} \ .
\label{nlnt}
\eeq
The inequality is used here because when one sets $x=-t$ and $y=-1/t$, some
$\lambda_{G,j}(x,y)$'s may become identically equal or may vanish.  There is
one case that we have considered where $N_{T,\lambda}=10$ but when one sets
$x=1/y=-t$, two of the $\lambda_{T,j}$'s become equal, so that
$N_{L,\lambda}=9$.  However, for all of the families discussed here,
(\ref{nlnt}) is realized as an equality.  A term $\lambda_{L,j}$ is dominant if
its magnitude $|\lambda_{L,j}|$ is greater than the magnitudes of other terms
$\lambda_{L,j^\prime}$ occurring in (\ref{vlsum}).  This definition is
motivated by the fact that in a given region of the complex $t$ plane, as $m$
gets large, the $\lambda_{L,j}$ with largest magnitude dominates the sum to a
greater and greater extent.  Thus, the continuous locus ${\cal B}$ where
$U_L(t)$ is nonanalytic occurs where there is a switching between dominant
terms $\lambda_{L,j}$ as one moves throughout the complex $t$ plane.  This
locus, i.e. the set of curves and possible line segments, arises as the
continuous accumulation set of the zeros of $V_L(t)$.  This is evident also in
the work of \cite{ww}.  There are also discrete points on the accumulation set,
at $t=e^{\pm 2 \pi i/3}$ \cite{ww}.

\section{The Family of Knots $A_n$ with $G_+(A_n)=D1C_{n-1}$}

\begin{figure}[hbtp]
\centering
\leavevmode
\epsfxsize=3.0in
\begin{center}
\leavevmode
\epsffile{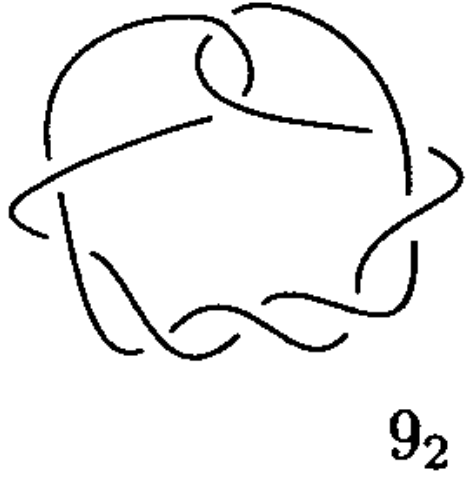}
\end{center}
\caption{\footnotesize{Example of the family $A_n$ for $n=9$.  This is the knot
$9_2$ in the standard labelling.}}
\label{k9_2}
\end{figure}

Let us define the (multi)graph $D1C_n$ as the result of doubling one edge
(denoted as $D1$) of the circuit graph $C_n$.  Note that $D1C_2=FL_3$.  These
multigraphs are associated with a certain family of knots.  We have found this
association by observing a number of special cases and then generalizing this
correspondence.  Specifically, we observe that
\beq
G_+(3_1)=D1C_2
\label{gp31}
\eeq
\beq
G_+(4_1)=D1C_3
\label{gp41}
\eeq
\beq
G_+(5_2)=D1C_4
\label{gp52}
\eeq
and
\beq
G_+(6_1)=D1C_5 \ .
\label{gp61}
\eeq
We generalize this to a family of knots with $r$ crossings,
denoted $A_n$ with the property that
\beq
G_+(A_n) = D1C_{n-1}
\label{gard}
\eeq
Thus, $A_3=3_1$, $A_4=4_1$, $A_5=5_2$, $A_6=6_1$, etc. 
As an example, we show in Fig. \ref{k9_2} a picture of the knot
$A_9$, which is $9_2$ in the standard list.  Higher members of this family are
obtained by adding more twists to the bottom part of the knot, keeping the
entwinement in the upper central part the same.

The knot $A_n$ has writhe 
\beq
w(A_n)=\cases{ -n & if $n$ is odd \cr
                4-n & if $n$ is even} \ .
\label{wan}
\eeq
The shaded diagram $D_+$ corresponding to this knot has 
\beq
n_d(D_+(A_n))=n-1
\label{ndan}
\eeq
dark regions and 
\beq
n_\ell(D_+(A_n))=3
\label{nellan}
\eeq
light regions.

We calculate the Jones polynomial for this family of knots by first calculating
the Tutte polynomial for the graph $D1C_n$.  We find 
\beqs
T(D1C_n,x,y) & = & (1+y)\Bigl [ y + \sum_{j=1}^{n-2}x^j \Bigr ] + x^{n-1} 
\cr\cr
& = & \frac{1}{(x-1)}\Bigl [ (1+y)(xy-x-y)+(x+y)x^{n-1} \Bigr ] \ .
\label{td1cn}
\eeqs
 Using the relation (\ref{vt}), we then have 
\beq
V_{A_n}(t) = \frac{t^k}{(1+t)}\biggl [ (1+t^{-2}) + 
(1-t^{-1})(1+t+t^{-1})(-t)^{1-n} \biggr ]
\label{van}
\eeq
where
\beq
k=\cases{ 0 & if $n$ is odd \cr
          3 & if $n$ is even} \ .
\label{k}
\eeq
This is of the form (\ref{vlsum}) with $N_{L,\lambda}=2$, the terms 
\beq
\lambda_{A_n,1}=1
\label{laman1}
\eeq
\beq
\lambda_{A_n,2}=-t^{-1}
\label{laman2}
\eeq
and the corresponding coefficients 
\beq
c_{A_n,1}=\frac{t^k(1+t^{-2})}{1+t}
\label{can1}
\eeq
\beq
c_{A_n,2}=\frac{t^k(1-t)(1+t+t^{-1})}{1+t} \ .
\label{can2}
\eeq
Our formula (\ref{van}) contains as special cases the known results for
\beq
V_{A_3}(t)=t^{-4}(-1+t+t^3)
\label{va3}
\eeq
\beq
V_{A_4}(t)=t^{-2}(1-t+t^2-t^3+t^4)
\label{va4}
\eeq
and so forth for higher $A_n$.

\begin{figure}[hbtp]
\centering
\leavevmode
\epsfxsize=3.0in
\begin{center}
\leavevmode
\epsffile{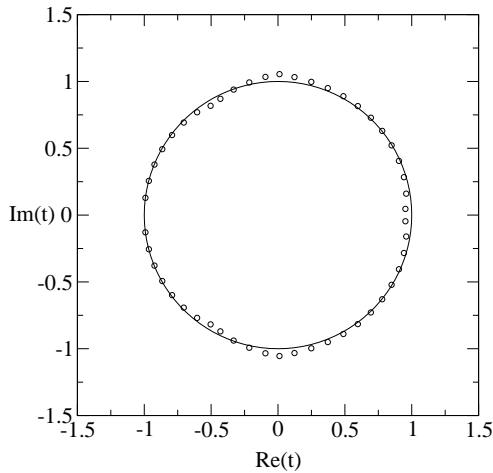}
\end{center}
\caption{\footnotesize{Singular locus ${\cal B}$ for the $n \to \infty$ limit
of the family $A_n$. For comparison, zeros of the Jones polynomial
$V_{A_n}(t)$ are shown for $n=50$.}}
\label{zan}
\end{figure}

In the limit $n \to \infty$, the continuous accumulation set ${\cal B}$ of the
zeros of $V_{A_n}(t)$ is the solution of the degeneracy equation $|t|=1$, i.e.,
the unit circle in the complex $t$ plane.  We find that the density of the
zeros on this circle is uniform.  Evidently, this locus separates the $t$ plane
into two regions.  Note that for this family, the discrete points given in
eq. (\ref{tpoints}) lie on the continuous accumulation set.  We denote the
region exterior to this circle, i.e., $|t| > 1$ as $R_1$ and the interior of
the circle, $|t| < 1$ as region $R_2$.  In region $R_1$, $\lambda_{A_n,1}$ is
dominant and
\beq
U(\{A\},t)=1 \quad {\rm for} \ \ t \in R_1 \ .
\label{uanr1}
\eeq
Thus, as $t \to \infty$, $U$ is a constant.  For other families of links, we
find that $U$ can increase in magnitude without bound as $t \to \infty$. 
In region $R_2$, $\lambda_{A_n,2}$ is dominant and 
\beq
|U(\{A\},t)|=\frac{1}{|t|}  \quad {\rm for} \ \ t \in R_2 \ .
\label{uanr2}
\eeq
$|U(\{A\},t)|$ diverges like $1/|t|$ as $t \to 0$.  Indeed, we find that for
each of the families of links that we have considered, $U$ diverges as $t \to
0$ (with the power depending on the family).  

\section{Families of Knots and Links $B_n$ with $G_+(B_n)=(Wh)_n$}

\begin{figure}[hbtp]
\centering
\leavevmode
\epsfxsize=3.0in
\begin{center}
\leavevmode
\epsffile{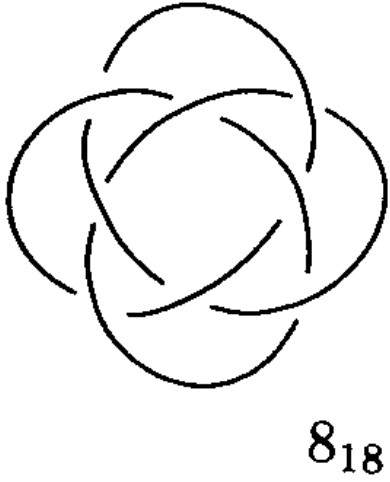}
\end{center}
\caption{\footnotesize{The knot $8_{18}$, as an example of the family 
$B_n$ for $n=5$.}}
\label{k8_18}
\end{figure}

We observe that a number of knots and links yield wheel graphs as their
associated graphs.  Specifically, we find that
\beq
G_+(4_1)=(Wh)_3=D1C_3
\label{gp41wh3}
\eeq
\beq
G_+(6^3_2)=(Wh)_4
\label{gp632}
\eeq
and
\beq
G_+(8_{18})=(Wh)_5
\label{gp818}
\eeq
A picture of the knot $8_{18}$ is shown in Fig. \ref{k8_18}. 
We generalize this to the family $B_n$ of links with 
$G_+(B_n)=(Wh)_n$.  The links in this family always have an even number of
crossings, $r(B_n)$.  This family has 
\beq
n(G_+(B_n)) \equiv n = \frac{r(B_n)}{2}+1
\label{nbr}
\eeq
\beq
n_d(D_+(B_n))=n_\ell(D_+(B_n))
\label{ndbn}
\eeq
and
\beq
w(B_n)=0 \ .
\label{wbn}
\eeq

For the Tutte polynomial of the wheel graph $(Wh)_n$ we find, in the notation
of (\ref{tgsum}), $N_\lambda=3$ and, with $m=n-1$,  
\beq
\lambda_{Wht,1}=1
\label{lam1wheel}
\eeq
\beq
\lambda_{Wht,(2,3)}=\frac{1}{2}\biggl [1+x+y \pm \Bigl [(1+x+y)^2-4xy \Bigr
]^{1/2} \biggr ]
\label{lam23wheel}
\eeq
\beq
c_{Wht,1}=q-2=xy-x-y-1
\label{c1wheel}
\eeq
and
\beq
c_{Wht,j}=1 \quad {\rm for} \ \ j=2,3
\label{c2wheel}
\eeq
so that 
\beq
T((Wh)_n,x,y)=xy-x-y-1 + (\lambda_{Wht,2})^{n-1}+(\lambda_{Wht,3})^{n-1} \ .
\label{twheel}
\eeq

Using the relation (\ref{vt}), we have
\beq
V_{B_n}(t)=t+t^{-1}+ (\lambda_{Wh,2})^{n-1}+(\lambda_{Wh,3})^{n-1}
\label{vbn}
\eeq
where
\beq
\lambda_{Wh,(2,3)}=\frac{1}{2}\biggl [1-t-t^{-1} \pm  \Bigl [-1-2(t+t^{-1})
+t^2+t^{-2} \Bigr ]^{1/2} \biggr ] \ .
\label{lam23t}
\eeq
Note the symmetry
\beq
V_{B_n}(t)=V_{B_n}(t^{-1})
\label{vbnsym}
\eeq
which follows from the self-dual property of the wheel graphs and the symmetry
(\ref{ttdual}) of the Tutte polynomial.  Our general formula subsumes the
following special cases for $B_3=4_1$, $B_5=8_{18}$:
\beq
V_{B_3}=t^{-2}(1-t+t^2-t^3+t^4)
\label{vb3}
\eeq
\beq
V_{B_5}=t^{-4}(1-4t+6t^2-7t^3+9t^4-7t^5+6t^6-4t^7+t^8) \ .
\label{vb5}
\eeq

\begin{figure}[hbtp]
\centering
\leavevmode
\epsfxsize=3.0in
\begin{center}
\leavevmode
\epsffile{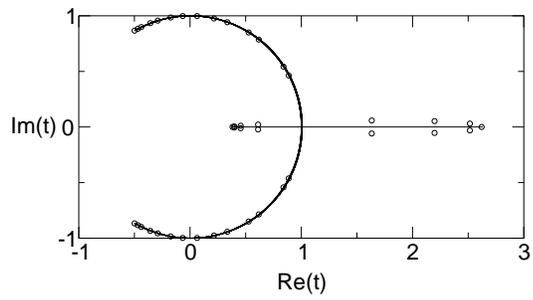}
\end{center}
\caption{\footnotesize{Singular locus ${\cal B}$ for the $n \to \infty$ limit
of the family $B_n$. For comparison, zeros of the Jones polynomial
$V_{B_n}(t)$ are shown for $n=42$.}}
\label{zbn}
\end{figure}

The locus ${\cal B}$ is determined by the degeneracy of leading $\lambda$'s, in
this case simply by $|\lambda_{Wh,2}|=|\lambda_{Wh,3}|$ and consists of the
union of an arc of the unit circle with a line segment on the positive real 
axis as shown in Fig. \ref{zbn}:
\beq
{\cal B}: \quad \{ \ t=e^{i\theta} \ , \quad -\frac{2\pi}{3} \le \theta \le 
\frac{2\pi}{3} \} \quad \bigcup \quad \{ \ \frac{3-\sqrt{5}}{2} \le t \le 
\frac{3+\sqrt{5}}{2} \ \} \ .
\label{bvb}
\eeq
Numerically, $(3-\sqrt{5})/2=0.381966...$ and $(3+\sqrt{5})/2=2.61803...$ The
four endpoints of ${\cal B}$ occur where the numerator of the function in the
square root, $1-2t-t^2-2t^3+t^4=(t^2+t+1)(t^2-3t+1)$ vanishes.  Note that the
endpoints of the circular arc occur precisely at the special points in
eq. (\ref{tpoints}). Evidently, the locus ${\cal B}$ does not separate the $t$
plane into different regions.  On the positive real axis for 
$t > (3+\sqrt{5} \ )/2$, the dominant term is $\lambda_{Wh,3}$ and hence
\beq
U(\{B\},t)=-\lambda_{Wh,3} \ .
\label{ub}
\eeq
This expression applies throughout the complex $t$ plane, taking account of the
fact that the sign of the square root reverses when one passes across the
branch cut associated with this square root, which is the same here as the 
locus ${\cal B}$.  Thus, in the vicinity of the origin, the dominant term is
$\lambda_{Wh,3}$ and $U(\{B\},t)$ diverges like $t^{-1}$ as $t \to 0$. As $t
\to \infty$, $U$ again diverges, like $t$. 

\section{Family of Links $E_n$ with $G_+(E_n)=H_{3,n}$}

\begin{figure}[hbtp]
\centering
\leavevmode
\epsfxsize=3.0in
\begin{center}
\leavevmode
\epsffile{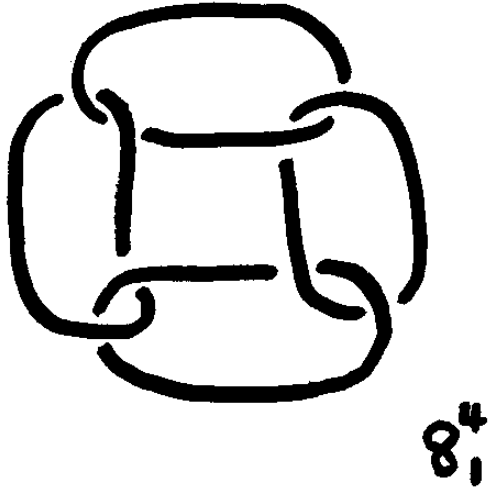}
\end{center}
\caption{\footnotesize{The knot $8^4_1$ as an example of the family 
$E_n$ for $n=4$.}}
\label{k8r4_1}
\end{figure}

In earlier work on continuous accumulation sets ${\cal B}$ of zeros of
chromatic polynomials as the number of vertices of a graph in some family goes
to infinity, a certain family of graphs called ``hammock'' graphs were studied
\cite{wa3} and it was shown that the locus ${\cal B}$ for these graphs was
noncompact in the $q$ plane, i.e. passed through the origin of the $1/q$ plane.
Thus, the zeros of the chromatic polynomials for these graphs have magnitudes
that are unbounded as one takes the number of vertices to infinity
\cite{wa3,sbounds}.  The hammock graph $H_{k,\ell}$ is defined as follows:
start with two end vertices and connect these together by $\ell$ ``ropes'',
each consisting of $k$ vertices, counting the end vertices.  The dual of the
graph $H_{3,\ell}$ is a multigraph which we denote $DC_\ell$:
\beq
(H_{3,\ell})^*=DC_\ell \ .
\label{dcnhammock}
\eeq
Here, $DC_\ell$ is the multigraph obtained by starting with a circuit graph and
doubling each edge.  Note that $DC_2=FL_4$ and $FL_4^*=C_4=H_{3,2}$. 
We calculate the correspondences
\beq
G_+(4^2_1)=DC_2 \quad , \ i.e., \ G_-(4^2_1)=H_{3,2}
\label{e2}
\eeq
\beq
G_+(6^3_1)=H_{3,3} \quad , \ i.e., \ G_-(6^3_1)=DC_3
\label{e3}
\eeq
\beq
G_+(8^4_1)=H_{3,4} \quad , \ i.e., \ G_-(8^4_1)=DC_4 \ .
\label{e4}
\eeq
We denote $E_2=R(4^2_1)$, $E_3=6^3_1$, and $E_4=8^4_1$ and generalize this to 
the family of links $E_n$ with the property
\beq
G_+(E_n)=H_{3,n} \quad , \ i.e., \ G_-(E_n)=DC_n \ .
\label{en}
\eeq
The links in this family always have an even number $r(E_n)$ of crossings.  
We note that the number of vertices in the associated graph, $|V(G_+(E_n))|$,
is given by 
\beq
|V(G_+(E_n))|=\frac{r(E_n)}{2}+2 = n+2  \ . 
\label{ner}
\eeq
In Fig. \ref{k8r4_1} we show a picture of the knot $8^4_1$ as an example of 
this family.

The family $E_n$ has
\beq
n_d(D_+(E_n))=n+2
\label{nden}
\eeq
\beq
n_\ell(D_+(E_n))=n \ .
\label{nellen}
\eeq
For a link containing $n_c \ge 2$ intertwined components (in contrast to a
knot), the writhe depends on the orientations chosen for these components
(circles) and we follow the convention that for a given link $L$, $[L]_1$
refers to the choice of orientations with the smallest value of the writhe,
$[L]_2$ to the next smallest, and so forth.  With an appropriate convention, we
have
\beq
w([E_n]_1)=-2n \ .
\label{wen}
\eeq

For the Tutte polynomial for the hammock graph $H_{3,n}$ we find
$N_{T,\lambda}=2$ and 
\beq
T(H_{3,n},x,y) = \frac{1}{(y-1)}\biggl [ (xy-x-y)(1+x)^n + (x+y)^n \biggr
] \ .
\label{tham}
\eeq
Using the relation (\ref{vt}), we find 
\beq
V_{E_n}(t)= -\frac{t^{1/2}}{1+t}\Biggl [ (1+t+t^{-1})\biggl ( 
\frac{1-t}{t^{3/2}} \biggr )^n + (-1)^n\biggl ( \frac{t+t^{-1}}{t^{3/2}}
\biggr )^n \Biggr ] \ .
\label{ven}
\eeq
This is of the form (\ref{vlsum}) with the terms 
\beq
\lambda_{E_n,1}=\frac{1}{\sqrt{t}}(-1+t^{-1})
\label{lam1en}
\eeq
\beq
\lambda_{E_n,2}=-\frac{1}{\sqrt{t}}(1+t^{-2})
\label{lam2en}
\eeq
and corresponding coefficients
\beq
c_{E_n,1}=-\frac{t^{1/2}(1+t+t^{-1})}{(1+t)}
\label{c1en}
\eeq
\beq
c_{E_n,2}=-\frac{t^{1/2}}{1+t} \ .
\label{c2en}
\eeq
This yields the special cases
\beq
V_{[R(4^2_1)]_1}(t)=t^{-9/2}(-1-t^2+t^3-t^4)
\label{v4212}
\eeq
\beq
V_{[6^3_1]_1}(t)=t^{-7}(1-t+3t^2-t^3+3t^4-2t^5+t^6) \ .
\label{v6311}
\eeq

\begin{figure}[hbtp]
\centering
\leavevmode
\epsfxsize=3.0in
\begin{center}
\leavevmode
\epsffile{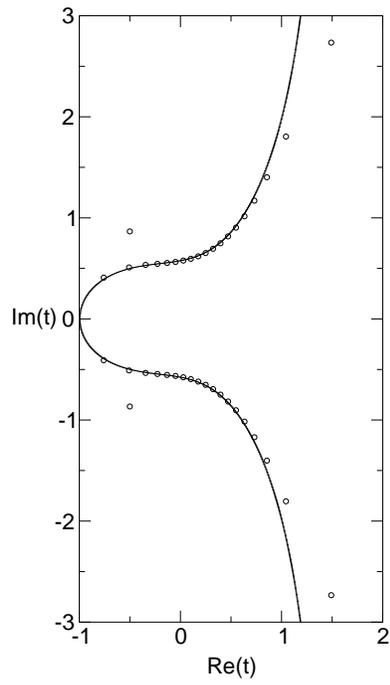}
\end{center}
\caption{\footnotesize{Singular locus ${\cal B}$ for the $n \to \infty$ limit
of the family $E_n$. For comparison, zeros of the Jones polynomial
$V_{E_n}(t)$ are shown for $n=42$.}}
\label{zen}
\end{figure}

The continuous locus ${\cal B}$ is determined by the degeneracy of magnitudes 
$|-1+t^{-1}|=|1+t^{-2}|$.  Let us introduce polar coordinates 
\beq
t=r e^{i\theta}
\label{tpolar}
\eeq
The locus ${\cal B}$ is described by the equation
\beq
-1 + \frac{1}{r^2} + 2r\cos\theta + 2\cos(2\theta) = 0
\label{polareq}
\eeq
i.e., 
\beq \cos\theta = \frac{1}{4} \biggl [ -r + \Bigl (
r^2+12-\frac{4}{r^2} \Bigr )^{1/2} \biggr ] \ .
\label{coseq}
\eeq
This curve passes through $t=-1$, crosses the imaginary axis at $t=\pm
\frac{i}{\sqrt{3}}$, passes into the $Re(t) > 0$ half-plane, then bends back
toward the imaginary axis and extends upward and downward to $\pm i \infty$.
Asymptotically,
\beq
\cos\theta = \frac{3}{2r} + O(\frac{1}{r^3}) \quad {\rm as} \ \ r \to
\infty \ .
\label{cosasymp}
\eeq
A portion of this curve is shown in Fig. \ref{zen}.  Thus, the locus ${\cal B}$
is noncompact in the $t$ plane, and this family of knots shows that the zeros
of Jones polynomials have magnitudes that are not necessarily bounded.  In
Fig. \ref{zen} we show ${\cal B}$ in the $t$ plane together with zeros
calculated for a link with $n=42$ crossings.  These zeros lie close to the
asymptotic locus ${\cal B}$ but, as expected for finite $n$, do not continue to
track near it as it extends arbitrary far from the origin.  This is similar to
the behavior that was found for the comparison of the location of chromatic
zeros and their asymptotic accumulation set for families of graphs where this
set was unbounded \cite{read91,w,wa,wa2,wa3,sbounds}.  Another noteworthy
feature of the locus ${\cal B}$ for this family is that it does not cross the
positive real $t$ axis, thereby showing that this is not a necessary property
of such a locus.  Evidently, the locus ${\cal B}$ divides the $t$ plane into
two different regions. For this family the special discrete points on the
accumulation set at the values given in eq. (\ref{tpoints}) are separate from
the continuous accumulation set ${\cal B}$.

In the respective regions $R_1$ and $R_2$ to the right and left of ${\cal B}$,
$\lambda_{E_n,2}$ and $\lambda_{E_n,1}$ are dominant, so that
\beq
U(\{E\},t)=-\lambda_{E_n,2} \quad {\rm for} \ \ t \in R_1
\label{uenr1}
\eeq
\beq
|U(\{E\},t)|=|\lambda_{E_n,1}| \quad {\rm for} \ \ t \in R_2
\label{uenr2}
\eeq
$U(\{E\},t)$ diverges like $t^{-5/2}$ as $t \to 0$ and vanishes like
$t^{-1/2}$ as $t \to \infty$. 

\begin{figure}[hbtp]
\centering
\leavevmode
\epsfxsize=3.0in
\begin{center}
\leavevmode
\epsffile{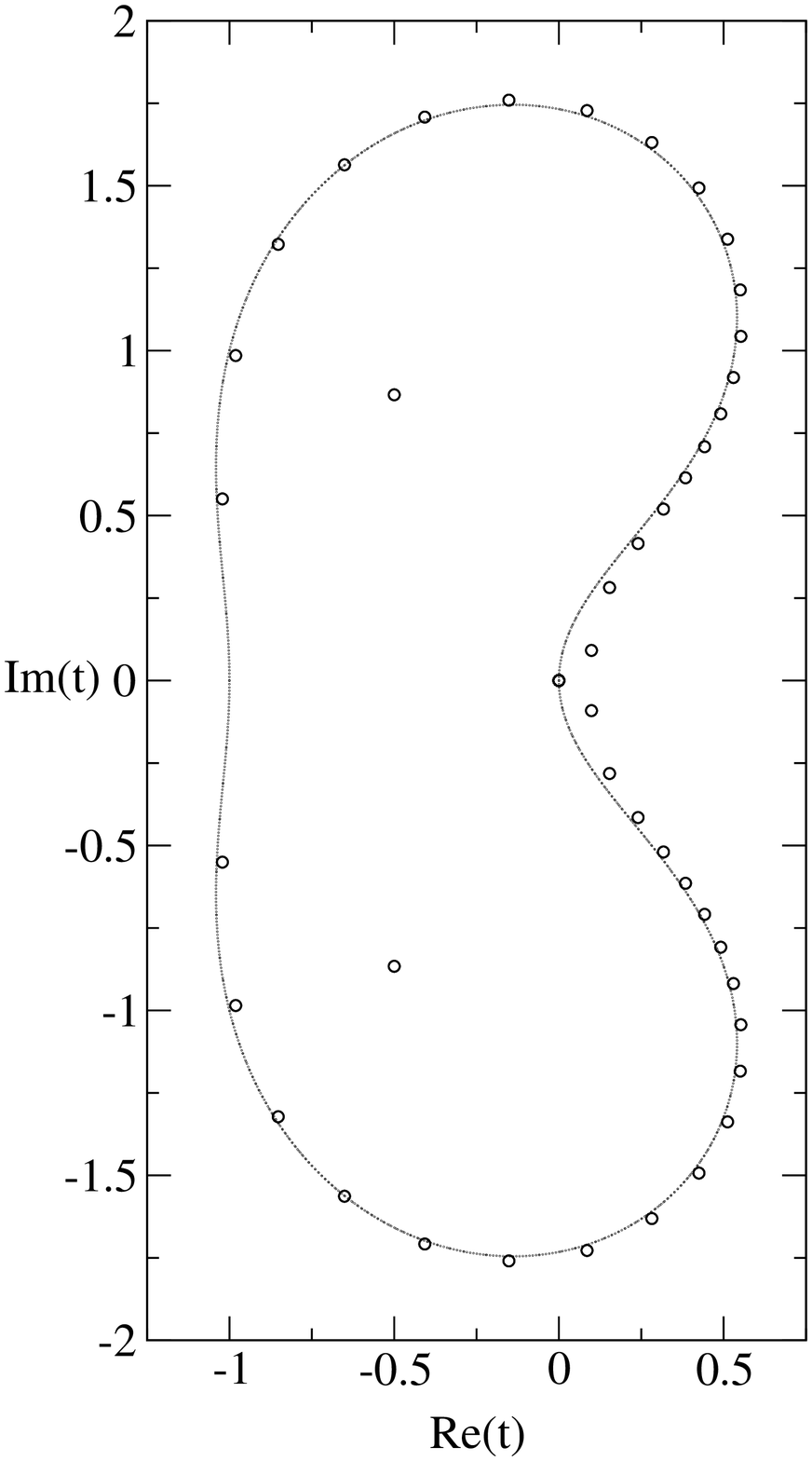}
\end{center}
\caption{\footnotesize{Singular locus ${\cal B}$ for the $n \to \infty$ limit
of the family $R(E_n)$ in the $t$ plane, or equivalently of the family $E_n$ in
the $1/t$ plane. For comparison, zeros of the Jones polynomial $V_{R(E_n)}(t)$
are shown for $n=42$.}}
\label{rzen}
\end{figure}

In the $s=1/t$ plane, the image of ${\cal B}$ is compact, forming a closed
curve passing through $s=-1$ and $s=0$.  Using the relation (\ref{vlvrl}), it
follows that the image of ${\cal B}$ in the $s$ plane is identical to the locus
${\cal B}$ for the family $R(\{E\})$ in the $t$ plane.  We show this in
Fig. \ref{rzen}.

\section{Families of Knots $F_n$ with $G(F_n)=HW_n$}

\begin{figure}[hbtp]
\centering
\leavevmode
\epsfxsize=3.0in
\begin{center}
\leavevmode
\epsffile{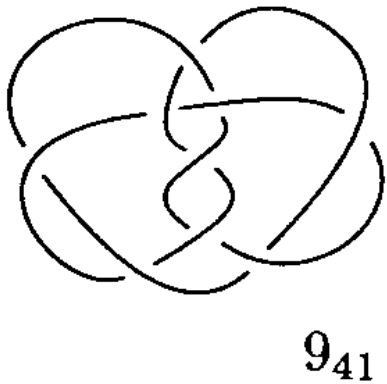}
\end{center}
\caption{\footnotesize{The knot $9_{41}$ as an example of the family $F_n$ for
$n=7$.}}
\label{k9_41}
\end{figure}

We define the graph $HW_n$ as the graph obtained by starting with the wheel
graph $(Wh)_m$ and performing a single homeomorphic expansion on each edge
forming a spoke of the wheel, where, as a result,
\beq
n=2m+1
\label{nmwh}
\eeq
is necessarily odd. 
We observe the correspondences $G_+(6_1)=HW_5$ and generalize this to a family
of knots $F_n$ with
\beq
G(F_n)=HW_n \ .
\label{gfn}
\eeq
Thus, the next member of the family beyond $F_5=6_1$ is $F_7=9_{41}$, and so
forth for higher members. An example of this family is shown in
Fig. \ref{k9_41}.

The knot $F_n$ has writhe 
\beq
w(F_n)=-\frac{(n-1)}{2} \ . 
\label{wfn}
\eeq
The shaded diagram $D_+$ corresponding to this knot has 
\beq
n_d(D_+(F_n))=n
\label{ndfn}
\eeq
dark regions and 
\beq
n_\ell(D_+(F_n))=\frac{n+1}{2} 
\label{nellfn}
\eeq
light regions.

We calculate the Jones polynomial for this family of knots by first calculating
the Tutte polynomial for the graph $HW_n$.  We find 
\beq
T(HW_n,x,y) = (xy-x-y-1)(x+1)^m + \lambda _{HW,1}^m + \lambda _{HW,2}^m
\label{thown}
\eeq
where $m=(n-1)/2$ as given by eq. (\ref{nmwh}) and 
\beq
\lambda_{HW,(1,2)}=\frac{1}{2}\biggl [1+2x+x^2+y \pm \Bigl
[(1+2x+x^2+y)^2-4x(x+1)(x+y) \Bigr
]^{1/2} \biggr ]
\label{lamhow12}
\eeq
Using the relation (\ref{vt}), we then have 
\beq
V_{F_n}(t) = 1 + (t+t^{-1})(1-t^{-1})^m + (t^{-2}-t^{-1}+1-t)^m 
\label{vfn}
\eeq
This is of the form (\ref{vlsum}) with $N_{L,\lambda}=3$, with the terms
\beq
\lambda_{F_n,1}=1
\label{lamfn1}
\eeq
\beq
\lambda_{F_n,2}=1-t^{-1}
\label{lamfn2}
\eeq
\beq
\lambda_{F_n,3}=t^{-2}-t^{-1}+1-t
\label{lamfn3}
\eeq
and the corresponding coefficients
\beq
c_{F_n,1}=1
\label{cfn1}
\eeq
\beq
c_{F_n,2}=t+t^{-1}
\label{cfn2}
\eeq
and
\beq
c_{F_n,3}=1\ .
\label{cfn3}
\eeq
Our formula (\ref{vfn}) contains as special cases the known results for
\beq
V_{F_5}(t)=t^{-4}(1-t+t^2-2t^3+2t^4-t^5+t^6)
\label{vf5}
\eeq
\beq
V_{F_7}(t)=t^{-6}(1-3t+5t^2-7t^3+8t^4-8t^5+8t^6-5t^7+3t^8-t^9)
\label{vf7}
\eeq
and so forth for higher $F_n$.

\begin{figure}[hbtp]
\centering
\leavevmode
\epsfxsize=3.0in
\begin{center}
\leavevmode
\epsffile{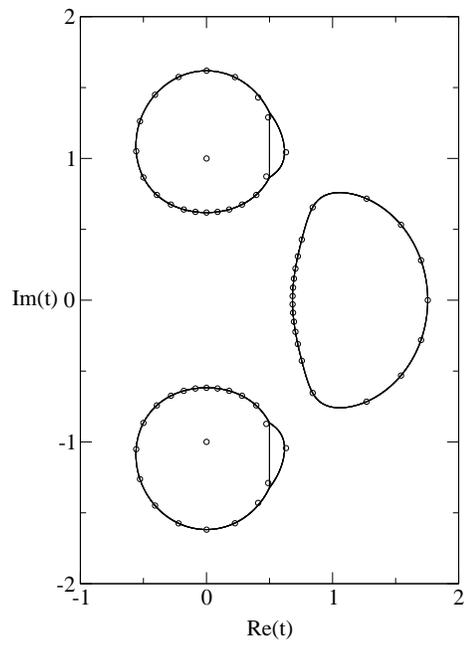}
\end{center}
\caption{\footnotesize{Singular locus ${\cal B}$ for the $n \to \infty$ limit
of the family $F_n$. For comparison, zeros of the Jones polynomial
$V_{F_n}(t)$ are shown for $n=63$.}}
\label{zfn}
\end{figure}

 From these results, we have then calculated the zeros of $V_{F_n}(t)$ and the
singular locus ${\cal B}$, which is shown in Fig.  \ref{zfn}.  This locus
divides the $t$ plane into six regions, of which two are self-conjugate and the
other four comprise two complex-conjugate pairs.  The two special points
(\ref{tpoints}) lie on two of the ovals comprising the locus ${\cal B}$. The
right-hand oval is the solution to the degeneracy equation $|\lambda_{F_n,1}|=
|\lambda_{F_n,3}|$; in the region $R_1$ outside of this oval, $\lambda_{F_n,3}$
is dominant, while inside the oval, in a region that we denote as $R_2$, 
$\lambda_{F_n,1}$ is dominant.  The large upper and lower ovals are, apart from
the small bubble regions, the solution to the equation 
$|\lambda_{F_n,2}|=|\lambda_{F_n,3}|$ and in the interior of these large
ovals,
which we call regions $R_3$ and $R_3^*$, $\lambda_{F_n,2}$ is dominant.
Finally, in the small bubble phases on the lower (upper) right part of 
the upper (lower) oval, which we denote $R_4$ and $R_4^*$, $\lambda_{F_n,1}$ is
dominant.  We thus have
\beq
U(\{F\},t) = (-\lambda_{F_n,3})^{1/2} \quad {\rm for} \ \ t \in R_1
\label{ufnr1}
\eeq 
\beq |U(\{F\},t)| = |\lambda_{F_n,1}|^{1/2}=1 \quad {\rm for} \ \ t \in
R_2, R_4, R_4^*
\label{ufnr2}
\eeq
\beq
|U(\{F\},t)| = |\lambda_{F_n,2}|^{1/2} \quad {\rm for} \ \ t \in R_3, R_3^*
\label{ufnr3}
\eeq
$|U(\{F\},t)|$ diverges like $|t|^{-1}$ as $t \to 0$. The locus ${\cal B}$
intersects the real axis at the two points $0.6823278...$ and $1.754877...$
(solutions of the cubic degeneracy equations given above).  As $t \to \infty$,
$U$ diverges like $t^{1/2}$.  Thus, the behavior of $U$ for large positive
real $t$ varies widely among the families of links that we have studied,
ranging from vanishing, to being a constant, to diverging.

\section{$V_L(t)$ for non-alternating links}  

For a non-alternating knot or link, one does not have a simple relation
between the Jones polynomial $V_L(t)$ and the Tutte polynomial as given in
eq. (\ref{vt}). In this case, there are two types of edges, $e_\pm(G)$, in
the associated graph instead of just one type of edge as in the case of an 
alternating link. In general, the Jones polynomial for any link can be 
expressed in term of Potts model partition function \cite{wu} as
\beq
V_L(t) = q^{-1/2}(-t^{-3/4})^{-w}Z(W_+,W_-)
\label{vz}
\eeq
where 
\beq
W_{\pm}(a,b) = A_{\pm} (1+v_{\pm}\delta _{ab})
\label{wpm}
\eeq
and 
\beq
v_{\pm} = e^{K_{\pm}} - 1 \ .
\label{vpm}
\eeq
The definition of $Z(W_+,W_-)$ related to the usual $Z(G,q,v_\pm)$ by
\beq
Z(W_+,W_-) = A_{+}^{e_+(G)}A_{-}^{e_-(G)} q^{-n(G)/2}Z(G,q,v_\pm)
\label{zz}
\eeq
where $n(G)$ is the total vertex umber of the associated graph. To satisfy
all three Reidemeister moves, we have to set $A_{\pm} = t^{\pm
1/4}$, $v_{\pm} = - (1+t^{\mp})$, and $q=t+2+1/t$. For alternating links,
all the edges are either $+$ type or $-$ type. Here we have $K_+ = - K_-$,
which means that we have both positive and negative spin-spin interaction 
with the same magnitude. The relation between $v_+$ and $v_-$ is that 
\beq
v_\pm = \frac{-v_\mp}{1+v_\mp}
\label{vpvm}
\eeq
Therefore, we have to modify eq. (\ref{cluster}) to be 
\beq
Z(G,q,v_-) = \sum_{G^\prime \subseteq G} q^{k(G^\prime)}v_-^{e(G^\prime)}
(\frac{-1}{1+v_-})^{e_+(G^\prime)}
\label{zvm}
\eeq
or 
\beq
Z(G,q,v_+) = \sum_{G^\prime \subseteq G} q^{k(G^\prime)}v_+^{e(G^\prime)}
(\frac{-1}{1+v_+})^{e_-(G^\prime)}
\label{zvp}    
\eeq
where $e_\pm(G^\prime)$ is the number of plus (minus) type edges in
$G^\prime$. Equivalently, we have 
\beq
T(G_{na},x,y) = \sum_{G^\prime \subseteq G} (x-1)^{k(G^\prime)-k(G)}
(y-1)^{c(G^\prime)} (\frac{-1}{y})^{e^\prime(G^\prime)}
\label{tpm}
\eeq
where $e^\prime(G^\prime)$ is the number of plus type edges in $G^\prime$
if $y=-t$, and the number of minus type edges in $G^\prime$ if $y=-1/t$.
Without loss of generality, consider the associated graph $G$ such that the
majority of its edges are plus type. If all vertices of $G$ are connected
by at least one of the plus edges, which can be achieved most of the time 
by performing the type III Reidemeister move, one can rewrite eq. 
(\ref{tpm}) as 
\beq
T(G_{na},x,y) = \sum_{\{e_-\}} \Delta T(G_{\{e_-\}},x,y) 
(\frac{-1}{y})^{e_-}
\label{tna}
\eeq
where we sum all possible minus type edge set $\{e_-\}$, and
$\Delta T(G_{\{e_-\}},x,y)$ is the difference of Tutte polynomials for the
graph with and without a particular edge set $\{e_-\}$. Combining eqs.
(\ref{ztutte}), (\ref{vz}), and (\ref{zz}), we have
\beqs
V(t)_{na} & = & (-t^{3/4})^w t^{(2-2n(G)+e_+(G)-e_-(G))/4}
T(G_{na},-t,-1/t) \cr\cr
 & = & (-t^{3/4})^w t^{(-2+2n(G)+e_+(G)-e_-(G))/4} T(G_{na},-1/t,-t) \ .
\label{vna}
\eeqs

Another way to deal with non-alternating knots or links is even simpler and
makes use of the skein relation for the Jones polynomial, 
\beq
\frac{1}{t} V_{L_+}(t) - t V_{L_-}(t) = \biggl [ \sqrt(t) -
\frac{1}{\sqrt(t)} \biggr ] V_{L_0}(t)
\label{skein}
\eeq
where the associated graph for $L_0$ corresponds to (i) the graph with a
contraction of the edge considered if the definition of $\pm$ crossings is the
same for the oriented links and that for the shading, and (ii) the graph with a
deletion operation performed on this edge if these two definitions are
different. The Jones polynomial can be obtained by using this relation to flip
the sign of the crossings until all the graphs become alternating, after which
one can apply eq. (\ref{vt}).

\section{A Remark on Computational Complexity}

The question of the computational complexity associated with determining basic
properties of knots and links, i.e., roughly speaking, the number of steps
needed to obtain this information, is a subject of considerable interest in
pure and applied mathematics and computer science (e.g., \cite{welsh}).
Recently, for example, an upper bound has been given for the number of
Reidemeister moves (RM) needed to transform the unknot to a circle with no
crossings, where the unknot is defined as the trivial knot \cite{hasslag}. This
bound is of the form $N_{RM} \le 2^{c_1 r}$, where, as before, $r$ is the
number of crossings and $c_1$ is a certain (large) constant.  We would like to
point out here that for the families of knots and links for which we have
calculated the Jones polynomial $V_L(t)$, our method, using the intermediate
step of computing the Tutte polynomial of the associated graph $G_+(L)$,
requires a polynomial, rather than exponential number of steps.  Specifically,
we use the standard iterative use of the deletion-contraction theorem for the
Tutte polynomial.  This difference in computational complexity is quite similar
to the situation for the Potts model partition function: a formal evaluation of
the sum over states for this model on an arbitrary graph involves exponentially
many steps since there are exponentially many spin configurations.  However, by
means of the transfer matrix method or the related iterated
deletion-contraction theorem for the equivalent Tutte polynomial, one can, for
specific families of graphs, carry out the evaluation in a polynomial number of
steps, since the partition function is just the trace of a power of the
transfer matrix.

Acknowledgment: We would like to thank Prof. F. Y. Wu for suggesting this
line of investigation and for giving us a copy of \cite{ww} prior to
publication.  This research was supported in part by the NSF grant PHY-9722101.

\vfill
\eject

\begin{thebibliography}{99}

\bibitem{rolfsen}{D. Rolfsen, {\it Knots and Links} (Publish or Perish,
Berkeley, 1976).} 

\bibitem{bz}{G. Burde and H. Zieschang, {\it Knots} (de Gruyter, New York,
1985).} 

\bibitem{y}{C. N. Yang, Phys. Rev. Lett. {\bf 19} (1967) 23.}

\bibitem{yg}{C. N. Yang and M.-L. Ge, {\it Braid Group, Knot Theory, and
Statistical Mechanics} (World Scientific, Singapore, 1988).}

\bibitem{kauffman}{L. Kauffman, {\it Knots and Physics} (World Scientific,
Singapore, 1991)}

\bibitem{wu}{F. Y. Wu, Rev. Mod. Phys. {\bf 64} (1992) 1099.}

\bibitem{welsh}{D. J. A. Welsh, {\it Complexity: Knots, Colourings, and
Counting} (Cambridge Univ. Press, Cambridge, 1993).} 

\bibitem{lick}{W. B. R. Lickorish, {\it An Introduction to Knot Theory}
(Springer, Berlin, 1997)}. 

\bibitem{boll}{B. Bollob\'as, {\it Modern Graph Theory} (Springer, New
York, 1998).}

\bibitem{jones85}{V. F. R. Jones, Bull. Am. Math. Soc. {\bf 12} (1985)
103.}

\bibitem{jones87}{V. F. R. Jones, Ann. Math. {\bf 126} (1987) 335.} 

\bibitem{jones89}{V. F. R. Jones, Pacific J. Math. {\bf 137} (1989) 311.} 

\bibitem{ww}{F. Y. Wu and J. Wang, Zeroes of the Jones polynomial, 
Northeastern preprint (Sept., 2000).} 

\bibitem{tutte1}{W. T. Tutte, Can. J. Math. {\bf 6} (1954) 80.}

\bibitem{tutte2}{W. T. Tutte, J. Combin. Theory {\bf 2} (1967) 301.}

\bibitem{tutte3}{W. T. Tutte {\it Graph Theory}, vol. 21 of
{\it Encyclopedia of Mathematics and its Applications}, ed. Rota,
G. C. (Addison-Wesley, New York, 1984).}

\bibitem{bbook}{N. L. Biggs, {\it Algebraic Graph Theory} (2nd ed., Cambridge
Univ. Press, Cambridge, 1993).}

\bibitem{w}{R. Shrock and S.-H. Tsai, Phys. Rev. {\bf E55} (1997) 5165.}

\bibitem{yl}{C. N. Yang and T. D. Lee, Phys. Rev.  {\bf 87} (1952) 404.}

\bibitem{ly}{T. D. Lee and C. N. Yang, Phys. Rev. {\bf 87} (1952) 410.}

\bibitem{fisher}{M. E. Fisher, in {\it Lectures in Theoretical Physics}
(Univ. of Colorado Press, Boulder, 1965), vol. 7C, pp. 1.}

\bibitem{ih}{V. Matveev and R. Shrock, J. Phys. A {\bf 28} (1995) 4859.}

\bibitem{yy}{V. Matveev and R. Shrock, Phys. Lett. {\bf A215} (1996) 271.}

\bibitem{wurev}{F. Y. Wu, Rev. Mod. Phys. {\bf 54} (1982) 235.}

\bibitem{kf}{P. W. Kasteleyn and C. M. Fortuin, J. Phys. Soc. Jpn. {\bf 26}
(Suppl.)  (1969) 11; C. M. Fortuin and P. W. Kasteleyn, Physica {\bf 57}
(1972) 536.}

\bibitem{rrev}{R. C. Read, J. Combin. Theory {\bf 4} (1968) 52.}

\bibitem{rtrev}{R. C. Read and W. T. Tutte, Chromatic Polynomials,
in {\it Selected Topics in Graph Theory, 3}, (Academic Press, New York,
1988), pp. 15.}

\bibitem{abe}{R. Abe, Prog. Theor. Phys. {\bf 38} (1967) 322.}

\bibitem{suzuki}{S. Ono, Y. Karaki, M. Suzuki, and C. Kawabata,
J. Phys. Soc. Jpn.  {\bf 25} (1968) 54.}

\bibitem{ipz}{C. Itzykson, R. Pearson, and J. B. Zuber, Nucl. Phys. {\bf B220}
(1983) 415.}

\bibitem{steph}{J. Stephenson and R. Couzens, Physica {\bf 129A} (1984) 201.}

\bibitem{saar}{W. van Saarloos and D. Kurtze, J. Phys. A {\bf 17} (1984) 1301.}

\bibitem{mm}{P. P. Martin and J. M. Maillard, J. Phys. A {\bf 19} L547 (1986).}

\bibitem{mbook}{P. P. Martin, {\it Potts Models and Related Problems in
Statistical Mechanics} (World Scientific, New York, 1991).} 

\bibitem{ms}{G. Marchesini and R. Shrock, Nucl. Phys. {\bf B318} (1989) 541. }

\bibitem{chisq}{V. Matveev and R. Shrock, J. Phys. A {\bf 28} (1995) 
1557, 5235.} 

\bibitem{ishs}{V. Matveev and R. Shrock, J. Phys. A. {\bf 28} (1995) L533.}

\bibitem{chw}{C. N. Chen, C. K. Hu, and F. Y. Wu, Phys. Rev. Lett. {\bf 76}
(1996), 169.}

\bibitem{wuetal}{F. Y. Wu, G. Rollet, H. Y. Huang, J. M. Maillard, C. K. Hu,
and C. N. Chen, Phys. Rev. Lett. {\bf 76} (1996) 173.}

\bibitem{pfef}{V. Matveev and R. Shrock, Phys. Rev. {\bf E54}, (1996)
6174.}

\bibitem{hcl}{H. Feldmann, R. Shrock, S.-H. Tsai, J. Phys. A (Lett.) {\bf 30}
(1997) L663.}

\bibitem{p}{H. Feldmann, R. Shrock, S.-H. Tsai, Phys. Rev. {\bf E57} (1998) 
1335.}

\bibitem{p2}{H. Feldmann, A. J. Guttmann, I. Jensen, R. Shrock, and S.-H. 
Tsai, J. Phys. A {\bf 31} (1998) 2287.}

\bibitem{bds}{N. L. Biggs, R. M. Damerell, and D. A. Sands, J. Combin. Theory
B {\bf 12} (1972), 123.}

\bibitem{bkw}{S. Beraha, J. Kahane, and N. N. Weiss, J. Combin. Theory B {\bf
27} (1979) 1.}

\bibitem{bkw80}{S. Beraha, J. Kahane, and N. N. Weiss, J. Combin. Theory B {\bf
28} (1980) 52.}

\bibitem{baxter}{R. J. Baxter, J. Phys. A {\bf 20} (1987) 5241.} 

\bibitem{read91}{R. C. Read and G. F. Royle, in {\it Graph Theory,
Combinatorics, and Applications} (Wiley, NY, 1991), vol. 2, p. 1009.}

\bibitem{ssbounds}{J. Salas and A. Sokal, J. Stat. Phys. {\bf 86} (1997)
551.}

\bibitem{strip}{M. Ro\v{c}ek, R. Shrock, and S.-H. Tsai, Physica
{\bf A252} (1998) 505.}

\bibitem{strip2}{M. Ro\v{c}ek, R. Shrock, and S.-H. Tsai, Physica 
{\bf A259} (1998) 367.}

\bibitem{hs}{R. Shrock and S.-H. Tsai, Physica {\bf A259} (1998) 315.}

\bibitem{wa}{R. Shrock and S.-H. Tsai, Phys. Rev. {\bf E56} (1997) 3935.}

\bibitem{wa2}{R. Shrock and S.-H. Tsai, Physica {\bf A265} (1999) 186.}

\bibitem{wa3}{R. Shrock and S.-H. Tsai, J. Phys. A {\bf 31} (1998) 9641.}

\bibitem{sbounds}{A. Sokal, Combin. Prob. Comput., to appear 
(cond-mat/9904146).}

\bibitem{wcyl}{R. Shrock and S.-H. Tsai, Phys. Rev. {\bf E60} (1999) 3512.}

\bibitem{wcy}{R. Shrock and S.-H. Tsai, J. Phys. A {\bf 32} (1999) 5053.}

\bibitem{pm}{R. Shrock, Phys. Lett. {\bf A261} (1999) 57.}

\bibitem{tk}{N. L. Biggs and R. Shrock, J. Phys. A {\bf 32} (1999) L489.} 

\bibitem{k}{S.-C. Chang and R. Shrock, Phys. Rev. {\bf E 62} (2000) 4650.}

\bibitem{s4}{S.-C. Chang and R. Shrock, Physica A {\bf 290} (2001) 402.}

\bibitem{tor4}{S.-C. Chang and R. Shrock, Physica A {\bf 292} (2001) 307.} 

\bibitem{t}{S.-C. Chang and R. Shrock, Ann. Phys. in press (cond-mat/0004129).}

\bibitem{sstran}{J. Salas and A. Sokal, cond-mat/0004330.}

\bibitem{js}{J. Jespersen and J. Salas, cond-mat/0011456.}

\bibitem{jsrs}{J. Salas and R. Shrock, cond-mat/0102190.}

\bibitem{bcc}{R. Shrock, in the {\it Proceedings of the 1999 British
Combinatorial Conference, BCC99}, Discrete Math., in press.} 

\bibitem{a}{R. Shrock, Physica A {\bf 283} (2000), 388.}

\bibitem{ta}{S.-C. Chang and R. Shrock, Physica A {\bf 286} (2000), 189.}

\bibitem{hca}{S.-C. Chang and R. Shrock, Physica A, in press
(cond-mat/0008477).} 

\bibitem{hasslag}{J. Hass and J. Lagarias, math.GT/9807012 (Jan., 2001).}

\end{thebibliography}
\end{document}